# Title: Imaging of Coronal Magnetic Field Reconnection leading to a Solar Flare

**Authors:** Yang Su[1*], Astrid Veronig[1], Gordon D. Holman[2], Brian R. Dennis[2], Tongjiang Wang[2,3], Manuela Temmer[1], Weiqun Gan[4]

**Affiliations:**

[1]Kanzelhöhe Observatory-IGAM, Institute of Physics, University of Graz, Universitaetsplatz 5/II, Graz 8010, Austria.

[2]Solar Physics Laboratory (Code 671), Heliophysics Science Division, NASA Goddard Space Flight Center, Greenbelt, MD 20771, USA.

[3]Department of Physics, the Catholic University of America, Washington, DC 20064, USA.

[4]Key Laboratory of Dark Matter and Space Astronomy, Purple Mountain Observatory, Chinese Academy of Sciences, Nanjing 210008, China.

[*]Correspondence and requests for materials should be addressed to Y. S. (yang.su@uni-graz.at).

**Abstract**:

Magnetic field reconnection is believed to play a fundamental role in magnetized plasma systems throughout the universe[1], including planetary magnetospheres, magnetars, and accretion discs around black holes. This letter presents extreme ultraviolet (EUV) and X-ray observations of a solar flare showing magnetic reconnection with a level of clarity not previously achieved. The multi-wavelength EUV observations from SDO/AIA show inflowing cool loops and newly formed, outflowing hot loops, as predicted. RHESSI X-ray spectra and images simultaneously show the appearance of plasma heated to >10 MK at the expected locations. These two data sets provide solid visual evidence of magnetic reconnection producing a solar flare, validating the basic physical mechanism of popular flare models. However, new features are also observed that need to be included in reconnection and flare studies, such as 3D non-uniform, non-steady, and asymmetric evolution.

**Main Text:**

The early concept of magnetic reconnection was proposed in the 1940s[1] to explain energy release in solar flares, the most powerful explosive phenomena in the solar system. The reconnection process reconfigures the field topology and converts magnetic energy to thermal energy, mass motions, and particle acceleration. The theories and related numerical simulations, especially 3D modelling, are still subjects of extensive research to obtain a full understanding of the process under different conditions. Meanwhile, observational studies have made progress in finding evidence of reconnection and deriving its physical properties to constrain and improve the theories.

In-situ measurements of the magnetic field, plasma parameters, and particle distributions have shown the existence of magnetic reconnection in laboratory plasmas[2,3], fusion facilities, and magnetospheres of planets[4,5]. Such in-situ measurements are still not possible in the extremely hot solar atmosphere. Instead, observations are obtained through remote sensing of emissions across the entire electromagnetic spectrum from radio to X rays and gamma rays. However, in the corona[6] the magnetic field pressure dominates the plasma pressure (low plasma beta) and the magnetic flux is "frozen into" the highly conductive plasma. As a result, the emitting plasma trapped in coronal loops outlines the geometry of the magnetic field and their structural changes reflect the changes of the field connectivity (in general). Considerable pieces of evidence for features likely linked to reconnection in solar flares[7,8] and coronal mass ejections (CMEs[9]) have been obtained so far. These include signatures of plasma inflow/outflow[10-14], hot cusp structures[15], current sheets[16-18], fast-mode standing shocks[19], and plasmoid ejection[20]. However, most evidence has been indirect and fragmented. Detailed observations of the complete picture are still missing due to the highly dynamic flare/CME process and limited observational capabilities.

The launch of the *Solar Dynamic Observatory* (SDO[21]) in 2010 significantly improved this situation. In particular, the Atmospheric Imaging Assembly (AIA[22]) has enabled continuous imaging of the full Sun in ten EUV, UV, and visible channels, with a spatial resolution of ~0.6 arc sec and a cadence of 12 s. Simultaneously, the *Ramaty High Energy Solar Spectroscopic Imager* (RHESSI[23]) is continuing to provide X-ray imaging and spectroscopic diagnostics of the heated plasma and accelerated electrons. The flare of interest was observed close to the southeast limb of the solar disk on 2011 August 17, but no CME was detected. The peak soft X-ray flux (1−8 Å) recorded by the X-ray monitor on a *Geostationary Operational Environmental Satellite* (GOES) was $2.3 \times 10^{-6}$ Watts m$^{-2}$, implying a modest flare magnitude of C2.3. The favourable timing, intensity, position, and orientation allowed us to observe the most complete evolution yet obtained of magnetic reconfiguration and energy release in an interacting coronal loop arcade.

The clearest visual evidence of the reconnection process comes from the synchronous imaging of cool inflow loops (Fig. 1a-b) and hot outflow loops (Fig. 1c, see also supplementary movie M1-M3). From ~04:05 UT to 04:28 UT, discrete coronal loops with temperatures from ~0.05 to 2 MK merged and disappeared near a central plane (the red line in Fig. 1b-c), the same location where a hot "X-shaped" structure (~10 MK, first image in Fig. 1c) gradually formed by ~04:10 UT. The two cusps above and below the X-type neutral point began to separate to form a "V-inverted V" structure at ~04:15 UT (second image in Fig. 1c). Newly formed hot loops (~10 MK) then appeared in two separated groups. Both flow away from the reconnection region (Fig. 1c), presumably due to magnetic tension forces.

The bright flare regions on the surface (observed at AIA 1600 Å, see Fig. 2b and supplementary Fig. 1) are the footpoints of coronal loops heated by thermal conduction and/or nonthermal electrons. As a result of the reconnection, the magnetic topology likely changed from an arcade of loops connecting footpoints A-B in the southern end and C-D in the northern end to two vertically separated new sets of loops (Fig. 2b). One group connects the lower cusp to regions A and C. This relaxes downward onto the flare arcade. The other group likely connected the higher cusp to the regions B and D. This expands outward, contributing to the loops (magnetic flux rope) building above the arcade.

X-ray emissions from the heated plasma confirm the energy release from a reconnection process in both timing and location. The RHESSI X-ray flux began to increase at 04:06:40 UT, close to the time when the EUV inflow loops visibly started moving together (Fig. 3d). The GOES 1-8 Å flux started to decrease after ~04:29 UT, the time when the last inflow loops (Fig. 3d) were seen disappearing. On the other hand, X-ray images and spectra show heated plasma (up to 17 MK, Fig. 2c) in the reconnected flare loops. The RHESSI coronal source in the 4-10 keV energy range (first and third images of Fig. 2a) indicates the presence of plasma at > 6 MK at/near the reconnection site. After 04:19 UT, a high coronal source appeared in the 10–20 keV images above the higher, relaxing cusp (Fig. 2a and Fig. 4b). This source and the extended source above the flare loops are known as double coronal X-ray sources, first observed in RHESSI images alone for a different flare[16]. Unlike the famous single Masuda source[19], which may be a signature of fast-mode standing shocks below the reconnection site, double coronal sources are thought to be related to heating in both the upper and lower outflow regions. With the support from the EUV images presented here, the double coronal sources become strong evidence for plasma heating following a reconnection process.

The formation of either the Masuda source or the double corona sources remains to be a question. In particular, the higher coronal source has been rarely detected and discussed[24]. The absence of a clear power-law bremsstrahlung component in the RHESSI spectra may indicate a negligible contribution of heating from non-thermal electrons in this case. One mechanism that could explain the double coronal sources is the slow mode magnetosonic shocks during flux retraction, which cause heating and density enhancements in the contracting loops ejected from the reconnection site[25].

In order to quantitatively investigate the inflow and outflow motions evident in the EUV movies, we defined two curves labelled C1 and C2 in Fig. 1c. The resulting time-distance plots for C1 (Figs. 3a) show coronal slices of loops that moved inward (toward the central line) from both sides and disappeared. The apparent inflow velocities increased from ~10 to over 50 km s$^{-1}$ as the loops approached the point of disappearance. The final inflow velocities ($V_{in}$) range from ~20 to ~70 km s$^{-1}$. The outflows are evident in the time-distance plots made for C2 from the 131 Å images (Figs. 3b). Signatures of some superposed fast-moving structures can be seen from 04:08 to 04:22 UT (solid cyan lines in Fig. 3d) along the paths of the two separating cusps (the two dashed cyan lines in Fig. 3d). We suggest that these structures are the new loops ejected from the reconnection region, while the two cusps are the subsequently heated thermal sources. The initial outflow velocities ($V_{out}$) range from ~90 to ~440 km s$^{-1}$. If we neglect the projection effect and assume[14] the measured outflow velocity is approximately the local Alfvén velocity ($V_A$), we estimate that the reconnection rate, $M_A = V_{in}/V_A \cong V_{in}/V_{out}$, varies from ~0.05 to 0.5 (Fig. 3e).

The event also revealed other new features of inflows. (1) Some inflow loops originated far from the reconnection site (up to 29,000 km). (2) The inflow loops, visible in different AIA channels, covered a wide range of temperatures from ~0.05 to 2 MK. This means that the plasma across the inflow field had different physical states. Compared with the steady, uniform case, the variable inflowing plasma could result in a variable heating rate and may help explain spikes observed in hard X-ray light curves during the impulsive phase of most flares. (3) The highly curved inflow loops expanded as they approached the reconnection site. Detailed flow velocity maps (Fig. 4b) derived from the Fourier Local Correlation Tracking (FLCT[26]) quantify this expansion. The diverging inflows, together with the signature of piling-up loops (white arrows in Fig. 3 and supplementary Fig. 2) and the high reconnection rates are suggestive of flux-pile-up reconnection[1], rather than the classic Petschek-like reconnection[27]. (4) The inflows were apparently asymmetric, assuming that inflow velocity vectors in both inflow regions are in the same plane. For example, the apparent inflow velocities at 04:20–21 UT were ~50 km s$^{-1}$ from the south and ~20 km s$^{-1}$ from the north. Since the reconnection process consumes same amount of opposite magnetic fluxes $|V_{in\_S}|B_S = |V_{in\_N}|B_N$ from the two inflow regions, one would expect that a weaker field requires a faster inflow speed. Then the region to the south would thus have a weaker field.

The inflow of magnetic loops is not a feature of the standard flare arcade reconnection model, in which loops within a linear arcade reconnect with each other. This inflow[13] and the lateral development of the flare arcade are intrinsically three-dimensional phenomena[28]. Realistic 3D simulations are required to understand the coronal magnetic reconnection and energy release processes observed here.

**Acknowledgments:**

The authors dedicate this paper to the late RHESSI PI, Robert P. Lin, in acknowledgement of his inspirational efforts that made possible the high quality solar X-ray data used in this paper. RHESSI is a NASA Small Explorer Mission. The Geostationary Operational Environmental Satellite Program (GOES) is a joint effort of NASA and the National Oceanic and Atmospheric Administration (NOAA). The Solar Dynamics Observatory (SDO) is a mission for NASA's Living With a Star (LWS) Program. The work of Y.S. and A.V. was supported by the European Community Framework Programme 7, High Energy Solar Physics data in Europe (HESPE), grant agreement No. 263086. Y. S. also acknowledges NSFC 11233008. The work of G. H. was supported by a NASA Guest Investigator Grant and the RHESSI program. The work of TW was supported by NASA grant NNX12AB34G and NASA Cooperative Agreement NNG11PL10A to CUA. M.T. acknowledges the Austrian Science Fund (FWF): V195-N16. W. G. acknowledges 2011CB811402 and NSFC 11233008.

Y. S. analysed the data, wrote the text and led the discussion. A. V., G. H., B. D., T. W., M. T. and W. G. contributed to the interpretation of the data and helped to improve the manuscript.

The authors declare no competing financial interests.

**Fig. 1. SDO/AIA overview of the reconnection process observed in EUV.** These false colour images are taken from the supplementary movies M1, M2 and M3 at different times during the flare on 2011 August 17. They show cool inflow loops merging horizontally (north/south) and hotter outflow loops separating vertically (east/west). The white curved line in each frame marks the visible edge (horizon) of the solar disk. The X and Y coordinates of each image are in arc seconds (1 arc sec ≈ 735 km) from disk center. (a) AIA images at 211, 193, 171 and 304 Å (1 Å = 0.1 nm) show inflowing loops with temperatures between ~0.05 and 2 MK. The white box in the third image indicates the field of view for Fig. 4. (b) Same as images shown in (a) after subtracting images taken one minute earlier. These difference images show moved/brightened inflow loops in white relative to their original positions/intensities in black. The red line in each frame marks the initial location of the X-structure where the inflow loops appeared to merge and disappear; the red arrows show the inflow directions. (c) AIA images at 131 Å showing plasma structures heated to ~10 MK. The yellow dashed curves marked C1 and C2 are used to derive the inflow and outflow profiles shown in Fig. 3. The two red arrows in the third image show the outflow directions of the hot, separating loops.

**Fig. 2. The reconnection and energy release imaged by SDO/AIA in EUV and RHESSI in X-rays.** The combination of data from different wavelengths shows the entire sequence of events expected for reconnection. (a) AIA 131 Å images, taken from supplementary movie M4, show the formation of the new hot loops. The images are rotated clockwise by 114º to orient the solar limb (black line) in the horizontal direction. RHESSI X-ray fluxes[30] are shown as red (blue) contours in the 4-10 (10-20) keV bands at 10, 40, and 80% of the maximum in each image. Bright flare regions on the surface observed at 1600 Å are superposed in pink in the fourth image. (b) The evolution of loop configurations. Orange lines show the major inflow loops derived from AIA images at 171, 193, 211 and 304 Å; cyan lines show the hot loops and cusps in the AIA 131 Å images. Main flare regions (pink) are marked by A, B, C, and D (see supplementary Fig. 1). The dashed grey lines in the fourth image illustrate the inflowing cool loops; the solid grey lines illustrate the newly formed loops after the reconnection. Arrows indicate flow directions. (c) RHESSI X-ray spectra of the heated plasma. The measured background-subtracted spectra (photons $s^{-1}$ $cm^{-2}$ $keV^{-1}$) are shown as histograms with error bars. The temperatures ($T_1$ and $T_2$) of two isothermal components (red and blue lines) that, summed together (black lines), give the best fit to the data are shown for each spectrum.

**Fig. 3. Time profiles of plasma inflow, plasma outflow, and X-ray flux.** (a) Time-distance plots (stack plots) showing (from the top) the time history of the intensity along curve C1 (the second image of Fig. 1c) at 171, 193, 211, and 304 Å. They are obtained from running ratio

difference images relative to the images taken 24 s earlier. The stack plots taken from the original images of these channels are shown in supplementary Fig. 2. The white arrow indicates the signature of piled-up loops. (b) Same as Fig. 3a but for curve C2 at 131 Å. (c) Flare light curves in RHESSI X-ray counts (3-6, 6-12, and 12-25 keV bands, left axis, color-coded curves) and GOES X-ray fluxes (1-8 Å, right axis, black curves). (d) Signatures of inflow (orange) and outflow (cyan) indicated in the stack plots shown above. The grey curve shows the time derivative (times $10^{10}$) of GOES 1-8 Å flux as a proxy of the flare heating rate[29]. The two dashed cyan lines show the locations of the two cusps with time. (e) Solid colour curves and plus signs show the calculated plasma inflow (orange) and outflow (cyan) velocities (left scale) as a function of time. Plus signs indicate that only a single velocity value was obtained. The black solid line shows the estimated reconnection rate ($M_A = V_{in}/V_A \cong V_{in}/V_{out}$) as a function of time.

**Fig. 4. Plasma flows and plasma heating in the reconnection region.** This figure shows the curved, expanding inflow loops and the heated plasma observed in X-rays. It supports the idea that plasma in the outflow regions are heated during flux retraction. The area covered by these plots is indicated by the white box in Fig. 1 and is rotated by 114° to match Fig. 2. (a) Red, green, and blue represent enhanced emission in the AIA 171, 193 and 131 Å images, respectively, relative to images taken 48 s earlier. (b) RHESSI 4 – 10 keV X-ray images with the colour code shown in the lower right expressed as the percentage of the peak intensity in each image. White contours in the fourth image show 10 – 20 keV X-ray intensity at 10, 20, 40, 60, and 90% of the peak value. The coloured arrows indicate the flow velocity vectors from 10 to 80 km s$^{-1}$ derived using the Fourier Local Correlation Tracking (FLCT[26]) method and a pair of AIA images taken 24 s apart in each of the same three channels. See also supplementary movie M5.

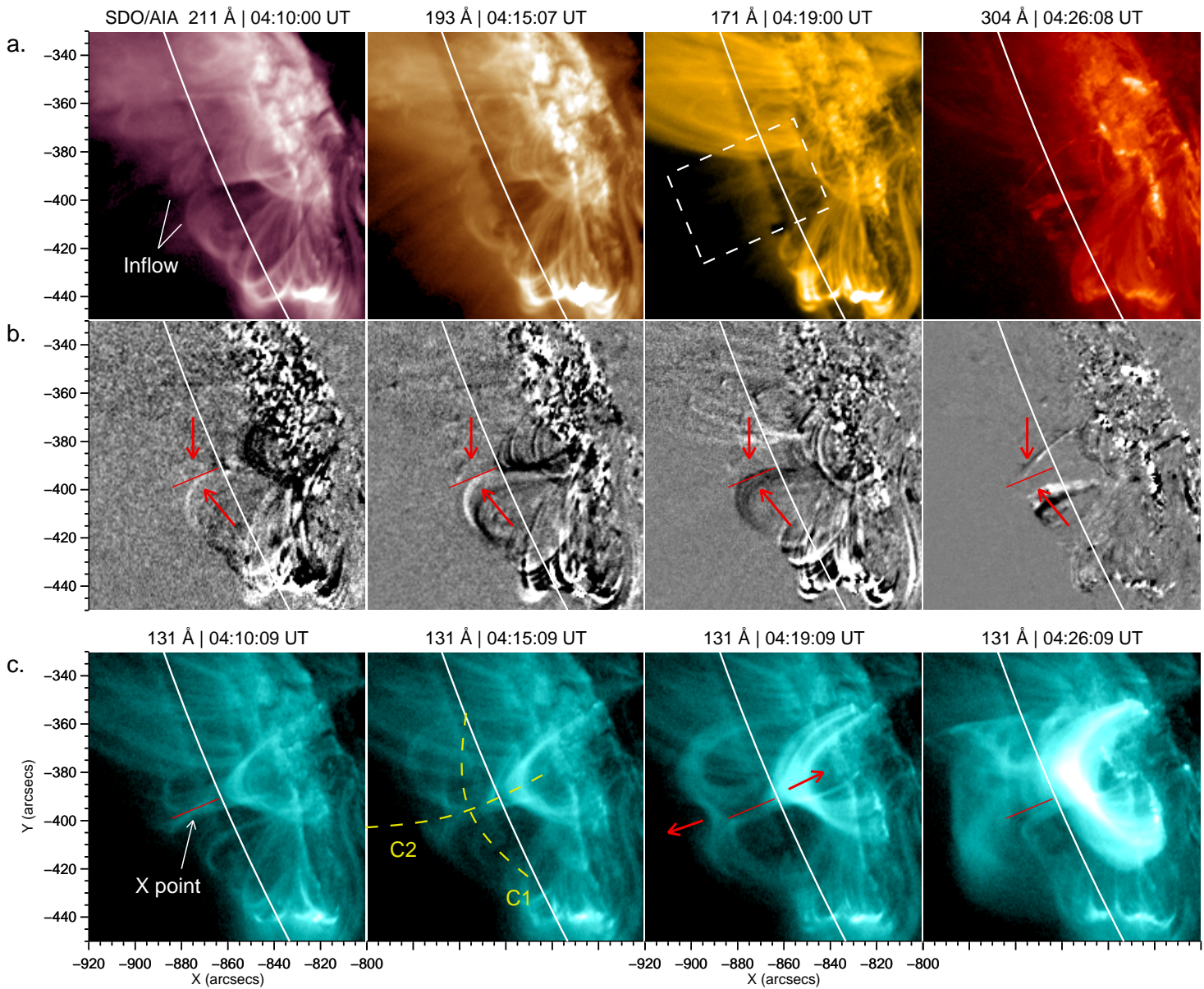

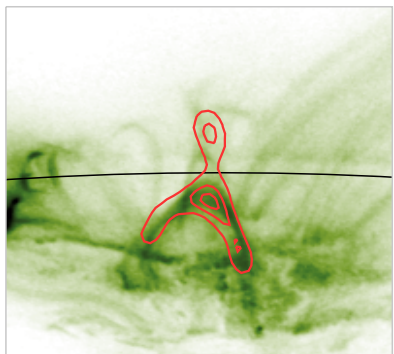 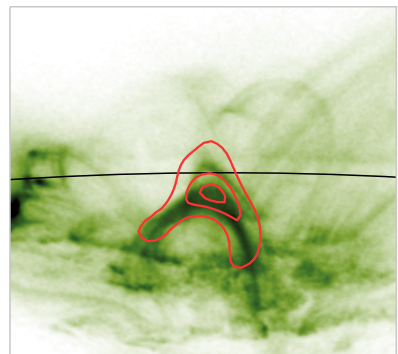 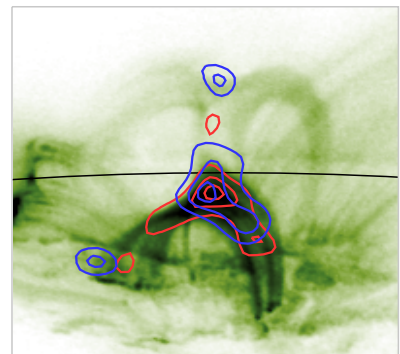 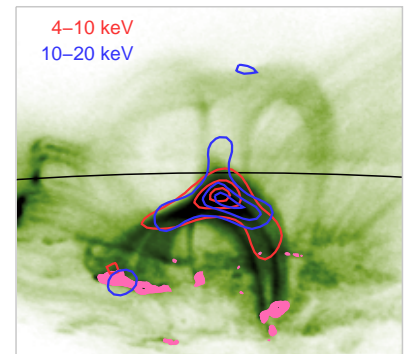
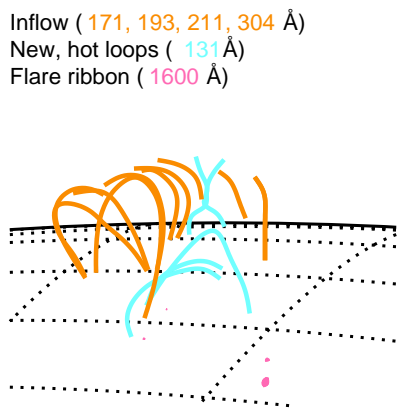 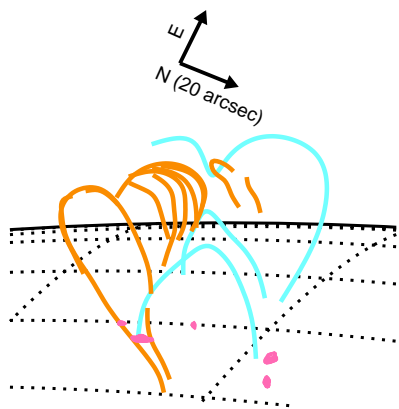 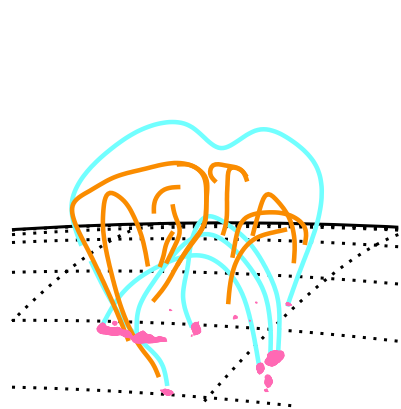 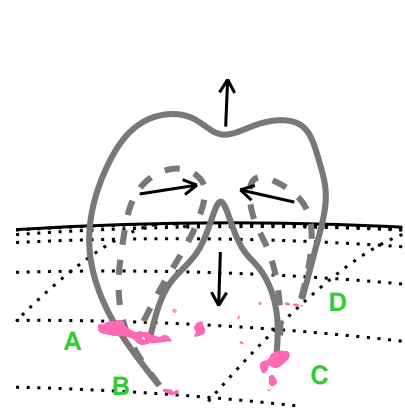
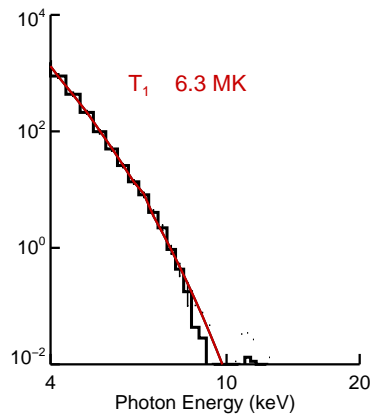 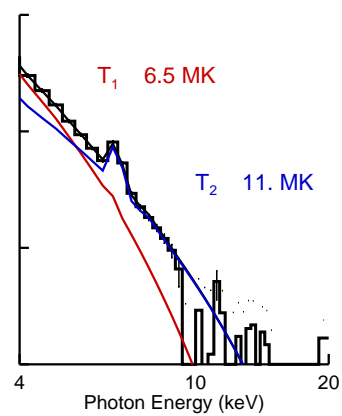 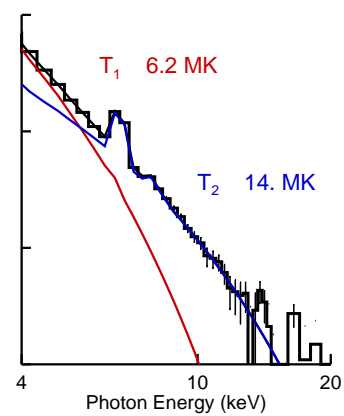 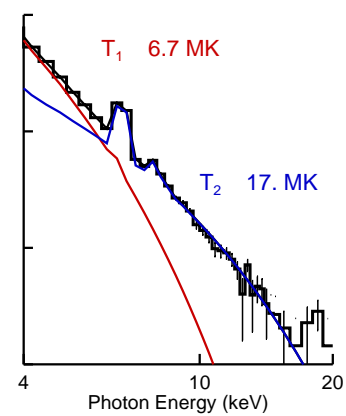

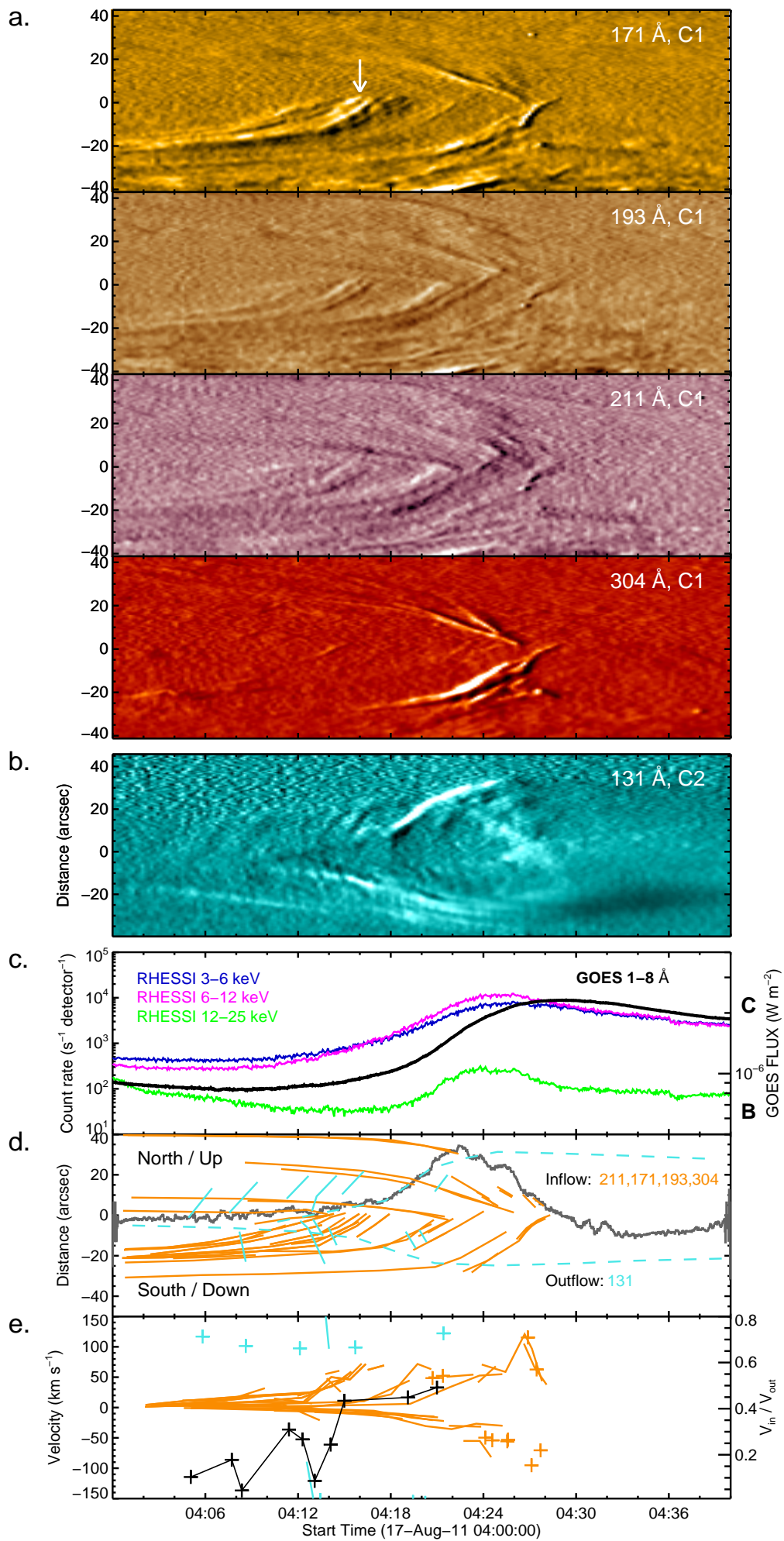

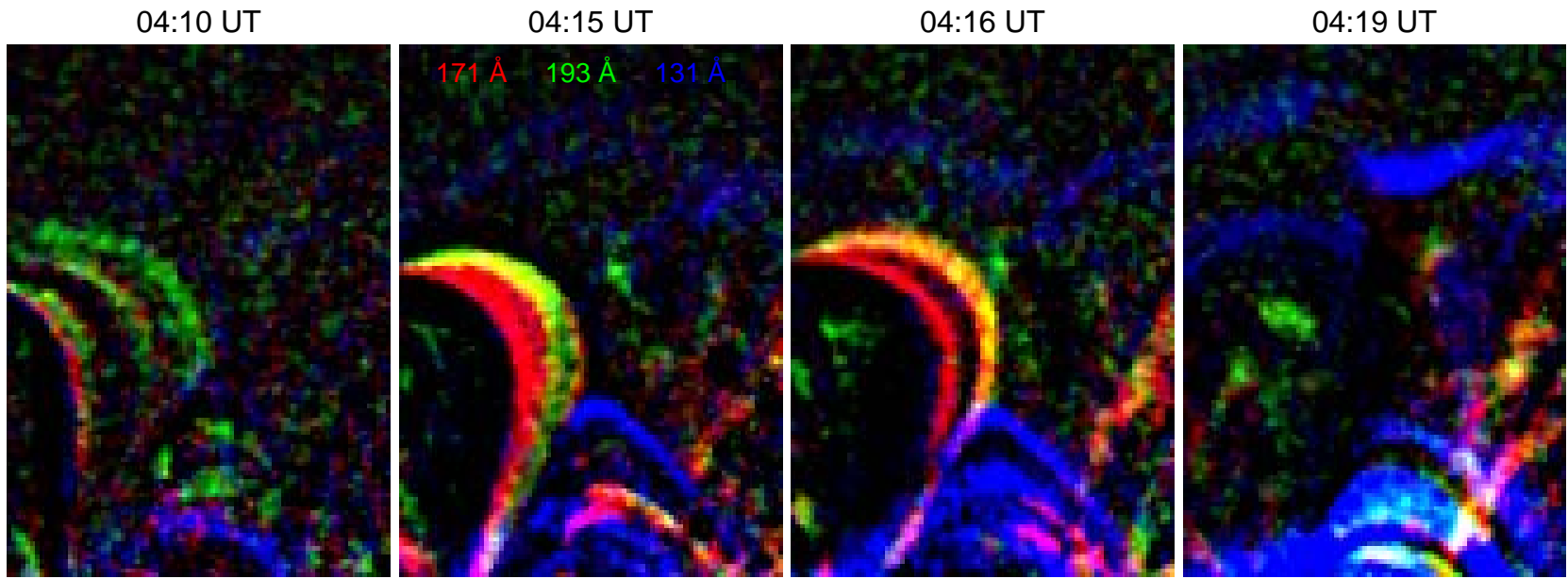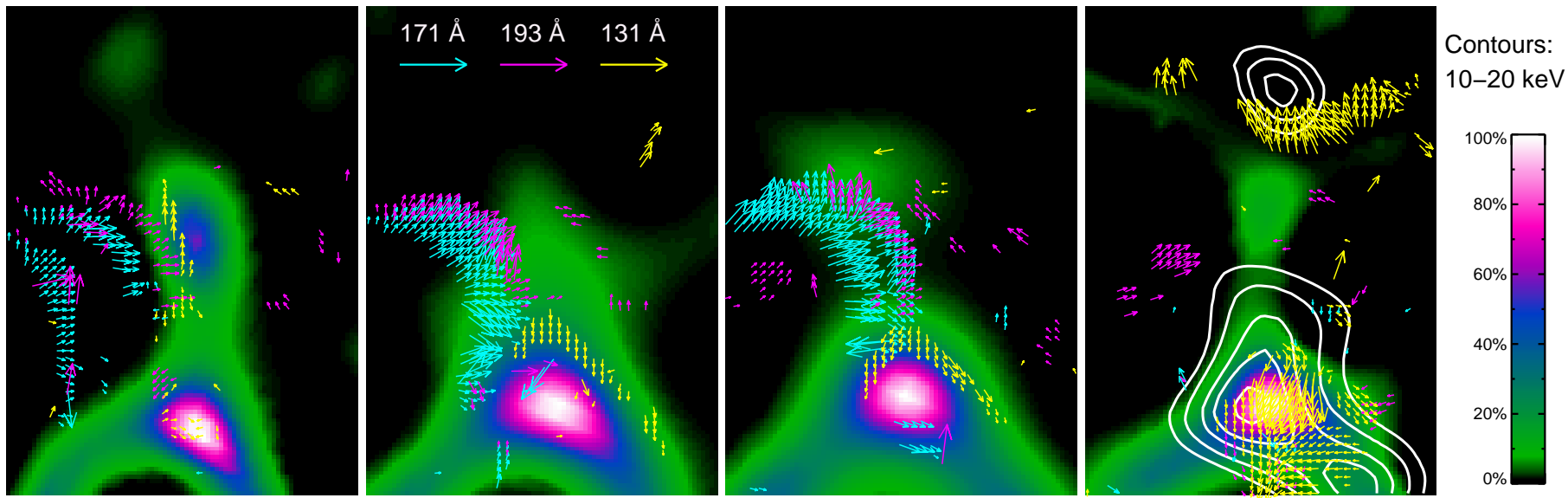